\documentclass[12pt,fleqn]{article}
\usepackage{amssymb}
\usepackage{graphics}
\usepackage[dvips]{color}
\newcommand{\vek}[1]{\mathrm{\bf #1}}

\newcommand{\mb}[1]{\mbox{\boldmath$#1$\unboldmath}}

\begin{document}
\begin{center}
\section*{Dimensionally hybrid Green's functions and density
of states for interfaces}
Rainer Dick\\[4mm]
Physics \& Engineering Physics, University of Saskatchewan,\\
116 Science Place, Saskatoon, SK S7N 5E2, Canada\\[5mm]
\end{center}

{\bf Abstract:}
The energy dependent Green's function $(E-H)^{-1}$ for an interface 
Hamiltonian which interpolates between two and three dimensions can be 
calculated explicitly. This yields an expression for the density of 
states $\varrho(E,z_0)$ on the interface which interpolates continuously 
between the two-dimensional $\varrho(E)=$constant behavior for high
energies and the three-dimensional $\varrho(E)\propto\sqrt{E}$ behavior 
for low energies.\\[5mm]
PACS:
05.30.Fk (Fermion systems and electron gas),
71.10.Pm (Electrons in reduced dimensions),
73.20.-r (Electron states at surfaces and interfaces).
%

\section{Introduction}\label{sec:intro}

Many quantities of physical interest depend on the number $d$ of spatial 
dimensions. This includes potentials and two-point correlation functions,
which are proportional to the zero energy Green's function
\begin{equation}\label{eq:greend}
G_{(d)}(r)=\left\{\begin{array}{cl}
-r/2,&\quad d=1,\\[2mm]
-(2\pi)^{-1}\ln(r/a),&\quad d=2,\\[2mm]
\Gamma\!\left(\frac{d-2}{2}\right)
\left(4\sqrt{\pi}^d r^{d-2}\right)^{-1},&
\quad d\ge 3,\\
\end{array}\right.
\end{equation}
as well as densities of states, which e.g. for non-relativistic free
particles are
\begin{equation}\label{eq:rhod}
\varrho_{(d)}(E)=g\Theta(E)\sqrt{\frac{m}{2\pi}}^d
\frac{\sqrt{E}^{d-2}}{\Gamma(d/2)\hbar^d}.
\end{equation}
These are densities of states per $d$-dimensional volume and per unit of 
energy. The factor $g$ is the number of helicity or spin 
states of the particles.\\ 
The corresponding dependence of the relation between the Fermi energy and
the density $n$ of electrons on $d$ is
\begin{equation}\label{eq:nfermid}
n_{(d)}=\frac{2}{\hbar^d\Gamma((d+2)/2)}\sqrt{\frac{mE_F}{2\pi}}^d.
\end{equation}

The generic case of physical interest is $d=3$, of course. 
Yet we frequently use mathematical techniques in $d=2$ for the
theoretical modeling of electrons or quasi-particles on surfaces or 
interfaces. But how two-dimensional is e.g. an ensemble of electrons in 
an interface? Is the two-dimensional logarithmic two-point correlation
function appropriate for the description of a gas of particles
in an interface? Or should we rather expect the three-dimensional
$r^{-1}$ correlation? The correct answer will certainly lie somewhere in
between, and how much in between will depend both on the specific system
and its parameters.\\
To address these kinds of questions analytically, dimensionally hybrid 
Hamiltonians of the form 
\begin{eqnarray}\label{eq:ham1}
H&=&\frac{\hbar^2}{2m}\int\!d^2\vek{x}\int\!dz\left(\nabla\psi^+\cdot
\nabla\psi+\partial_z\psi^+\cdot\partial_z\psi\right)
+\int\!d^2\vek{x}\int\!dz\,\psi^+V\psi
\\ \nonumber
&+&\left.\frac{\hbar^2}{2\mu}\int\!d^2\vek{x}\,\nabla\psi^+\cdot
\nabla\psi\right|_{z=0}
\end{eqnarray}
were introduced in \cite{rdtp03}. Here the convention is to use vector
notation $\vek{x}=(x,y)$, $\nabla=(\partial_x,\partial_y)$ for directions
parallel to the interface, while $z$ is orthogonal to the interface.\\
The potential term will generically also include three-dimensional and 
two-dimensional terms,
\[
V(\vek{x},z)=V_b(\vek{x},z)+V_i(\vek{x})\delta(z),
\]
but the competition between two-dimensional and three-dimensional 
behavior of physical quantities turns out to be mostly a consequence of 
competition between the two-dimensional and three-dimensional kinetic 
terms.\\
The two-dimensional mass parameter $\mu$ is a mass per length. In simple
models it is given by
\[
\mu=\frac{m}{L_\perp},
\]
where $L_\perp$ is a bulk penetration depth of states bound to the
interface at $z=0$, see Section \ref{sec:ham}.\\
The zero energy Green's function for the Hamiltonian
\begin{equation}\label{eq:ham0}
H_0=\frac{\hbar^2}{2m}\int\!d^2\vek{x}\int\!dz\left(\nabla\psi^+\cdot
\nabla\psi+\partial_z\psi^+\cdot\partial_z\psi\right)
+\left.\frac{\hbar^2}{2\mu}\int\!d^2\vek{x}\,\nabla\psi^+\cdot
\nabla\psi\right|_{z=0}
\end{equation}
for perturbations in the interface ($z'=0$, 
$G(\vek{x}-\vek{x}',z)=\langle\vek{x},z|G|\vek{x}',0\rangle$) 
satisfies
\[
(\Delta+\partial_z^2)G(\vek{x}-\vek{x}',z)+\frac{m}{\mu}\delta(z)\Delta
G(\vek{x}-\vek{x}',0)=-\delta(\vek{x}-\vek{x}')\delta(z)
\]
and was found in \cite{rdtp03} ($r=|\vek{x}-\vek{x}'|$),
\begin{equation}\label{eq:g03}
G(\vek{x}-\vek{x}',z)=\frac{1}{4\pi}\int_0^\infty\!dk\,
\frac{\exp(-k|z|)}{1+k\ell}J_0(kr),\quad 
\ell=\frac{m}{2\mu}.
\end{equation}
The Green's function in the interface is given in terms of a
Struve function and a Bessel function,
\begin{equation}\label{eq:g02}
G(\vek{x}-\vek{x}',0)=
\frac{1}{8\ell}\left[\mb{H}_0\!\left(\frac{r}{\ell}\right)
-Y_0\!\left(\frac{r}{\ell}\right)\right]
\end{equation}
and interpolates between two-dimensional and three-dimensional
distance laws,
\begin{eqnarray}\label{eq:rllell}
r\ll\ell:\quad
G(\vek{x}-\vek{x}',0)&=&\frac{1}{4\pi\ell}\left[-\gamma-
\ln\!\left(\frac{r}{2\ell}\right)+\frac{r}{\ell}
+\mathcal{O}\!\left(\frac{r^2}{\ell^2}\right)\right],
\\ \label{eq:rggell}
r\gg\ell:\quad
G(\vek{x}-\vek{x}',0)&=&\frac{1}{4\pi r}\left[1-\frac{\ell^2}{r^2}
+\mathcal{O}\!\left(\frac{\ell^4}{r^4}\right)\right].
\end{eqnarray}


\begin{center}
\begin{figure}[t]
\hspace*{50mm}\scalebox{0.4}{\includegraphics{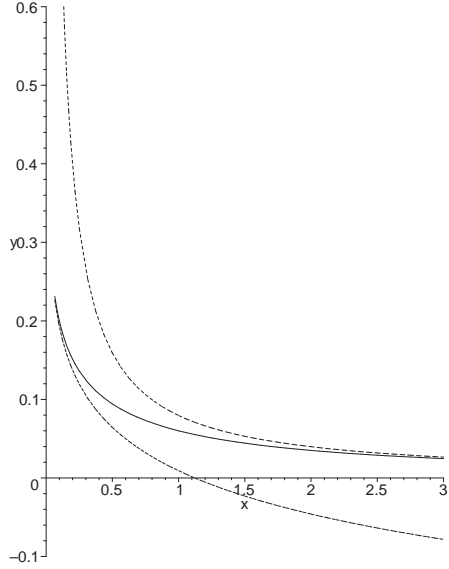}}
\caption{\label{fig:223pot}
The upper dotted line is the three-dimensional Green's function 
$(4\pi r)^{-1}$ in units of $\ell^{-1}$, the continuous line is the 
Green's function (\ref{eq:g02}), and the lower dotted line is the 
two-dimensional logarithmic Green's function. $x=r/\ell$.}
\end{figure}
\end{center}

 For the model (\ref{eq:ham0}) the Green's function $G(\vek{x},0)$ 
appears in the first place as a two-point correlation function in the 
interface, but it can also be realized as an electromagnetic potential 
for an electromagnetic Hamiltonian 
\begin{equation}\label{eq:23hamem}
H[F]=\frac{1}{2}\int\!d^2\vek{x}\int\!dz\left(
\vek{E}^2+E_\perp^2+\vek{B}^2+B_\perp^2\right)
+\left.\ell\int\!d^2\vek{x}\left(\vek{E}^2+B_\perp^2\right)
\right|_{z=0},
\end{equation}
if the fields which are continuous across the interface
yield a special interface contribution due to a finite limit
\[
2\ell=\lim_{L=0}\epsilon_r L.
\]
Here $\epsilon_r $ is the relative permittivity of the interface
and $L$ is its thickness.\\
The correspondence between electronic two-point correlation
functions and electrostatic potentials is often used to map
the quantum partition function of a free fermion gas into the
partition function of a classical Coulomb gas or plasma. The
two-dimensionial Coulomb gas also plays an important role in the 
variational treatment of the fractional quantum Hall effect 
\cite{rbl,TP,tapash}.
The appearance of $G(\vek{x},0)$ as the Green's function for both
of the dimensionally hybrid Hamiltonians (\ref{eq:ham0}) and
(\ref{eq:23hamem}) indicates that the free fermion gas -- Coulomb
gas duality persists in the transition regime between two and
three dimensions. However, the objective of the present paper is the 
extension of $G$ to non-zero energy and for perturbations off the
interface, and the discussion of implications 
on the density of states.\\
I will further elaborate on the motivation and justification for 
dimensionally hybrid Hamiltonians of the form (\ref{eq:ham1}) in 
Section \ref{sec:ham}. The calculation of the energy dependent
Green's functions and the resulting density of states on the interface
will be discussed in Sections \ref{sec:green} and \ref{sec:rho}.
Section \ref{sec:conc} contains the conclusions.

\section{Dimensionally hybrid Hamiltonians}\label{sec:ham}

The interesting part about the Hamiltonian (\ref{eq:ham1}) is the
competition between two-dimensional and three-dimensional kinetic 
terms. Hamiltonians with low-dimensional $\delta$ contributions
to the potential $V$ are standard in quantum mechanics, and we easily 
understand the model assumptions behind the use of low-dimensional 
potentials.
On the other hand, Hamiltonians with competing low-dimensional and 
three-dimensional kinetic terms are certainly not as common as 
superpositions of low-dimensional and three-dimensional potentials. I 
will denote Hamiltonians with superpositions of low-dimensional and 
three-dimensional kinetic terms as dimensionally hybrid Hamiltonians.
The introduction of these Hamiltonians in \cite{rdtp03} was based on 
physical intuition and mathematical curiosity. The primary physical 
justification for the mixed kinetic terms is electrons or quasi-particles
which can propagate with an effective mass $m_\ast$ in a thin 
layer of thickness $L$ and with mass $m$ outside of the layer.
In that case we would expect that particle propagation in the 
bulk-layer-bulk system should be described by a Hamiltonian of the form
(\ref{eq:ham1}) with $\mu\simeq m_\ast/L$.
The purpose of this section is to point out that models of the 
kind (\ref{eq:ham1}) can also be motivated in a different way from 
simple standard Hamiltonians with a standard three-dimensional kinetic 
term.\\
As the simplest possible model consider an interface
with an attractive potential
\[
V(\vek{x},z)=-W\delta(z),
\]
\[
H=\int\!d^2\vek{x}\int\!dz\left(\frac{\hbar^2}{2m}
\left(\nabla\psi^+\cdot\nabla\psi
+\partial_z\psi^+\cdot\partial_z\psi\right)+\psi^+ V\psi\right).
\]
The eigenvalues and eigenfunctions factorize with one set
of states bound to the interface,
\begin{equation}\label{eq:bound}
\psi_{\vek{k},\kappa,+}(\vek{x},z)
=\langle\vek{x},z|\vek{k},\kappa,+\rangle
=\frac{\exp(\mathrm{i}\vek{k}\cdot\vek{x})}{2\pi}
\frac{\sqrt{mW}}{\hbar}
\exp\!\left(-\frac{m}{\hbar^2}W|z|\right),
\end{equation}
\[
E=\frac{\hbar^2}{2m}\left(\vek{k}^2-\kappa^2\right)
=\frac{\hbar^2}{2m}\vek{k}^2-\frac{m}{2\hbar^2}W^2,
\quad
\kappa=\frac{mW}{\hbar^2},
\]
and two orthogonal sets of transversally free  states
with transversal wavenumber $k_\perp\ge 0$ and energy
$E=\hbar^2(\vek{k}^2+k^2_\perp)/2m$,
\begin{equation}\label{eq:free1}
\psi_{\vek{k},k_\perp,-}(\vek{x},z)
=\langle\vek{x},z|\vek{k},k_\perp,+\rangle
=\frac{\exp(\mathrm{i}\vek{k}\cdot\vek{x})}{2\pi}
\frac{\hbar^2 k_\perp\cos\!\left(k_\perp z\right)
-mW\sin\!\left(k_\perp |z|\right)}{
\sqrt{\pi\left(\hbar^4 k^2_\perp+m^2W^2\right)}},
\end{equation}
\begin{equation}\label{eq:free2}
\psi_{\vek{k},k_\perp,-}(\vek{x},z)
=\langle\vek{x},z|\vek{k},k_\perp,-\rangle
=\frac{\exp(\mathrm{i}\vek{k}\cdot\vek{x})}{2\pi}
\frac{1}{\sqrt{\pi}}\sin\!\left(k_\perp z\right).
\end{equation}

The completeness relation for the factor $\chi(z)$
in $\psi(\vek{x},z)=\phi(\vek{x})\chi(z)$ is \cite{patil}
\[
|\kappa,+\rangle\langle\kappa,+|+
\int_0^\infty\!dk_\perp\left(|k_\perp,+\rangle\langle k_\perp,+|
+|k_\perp,-\rangle\langle k_\perp,-|\right)=1.
\]
This implies a decomposition of the kinetic operator
\[
\frac{\hbar^2}{2m}\int\!d^2\vek{x}\int_{-\infty}^\infty\!dz
\left(\chi^+(z)\nabla\phi^+(\vek{x})\cdot\nabla\phi(\vek{x})
\chi(z)+\partial_z\chi^+(z)\phi^+(\vek{x})\cdot\phi(\vek{x})
\partial_z\chi(z)\right)
\]
\[
=\left.\frac{\hbar^2}{2m}\int\!d^2\vek{x}\int_{-\infty}^\infty\!dz
\left(\nabla\psi^+(\vek{x},z)\cdot\nabla\psi(\vek{x},z)
+\partial_z\psi^+(\vek{x},z)\cdot\partial_z\psi(\vek{x},z)\right)
\right|_{\mathrm{free}}
\]
\[
+\frac{\hbar^2}{2m}\int\!d^2\vek{x}\,
\nabla\phi^+(\vek{x})\cdot\nabla\phi(\vek{x})
+\frac{mW^2}{2\hbar^2}\int\!d^2\vek{x}\,\phi^+(\vek{x})\cdot\phi(\vek{x}),
\]
where the first kinetic term on the right hand side only acts on 
components which are unbound in $z$ direction, while the last two terms 
arise from the bound states.
If we now express both the operators acting on free states and the 
operators acting on bound states in terms of $\psi(\vek{x},z)$, we find
\[
\frac{\hbar^2}{2m}\int\!d^2\vek{x}\int_{-\infty}^\infty\!dz
\left(\nabla\psi^+(\vek{x},z)\cdot\nabla\psi(\vek{x},z)
+\partial_z\psi^+(\vek{x},z)\cdot\partial_z\psi(\vek{x},z)
\right)
\]
\[
=\left.\frac{\hbar^2}{2m}\int\!d^2\vek{x}\int_{-\infty}^\infty\!dz
\left(\nabla\psi^+(\vek{x},z)\cdot\nabla\psi(\vek{x},z)
+\partial_z\psi^+(\vek{x},z)\cdot\partial_z\psi(\vek{x},z)
\right)\right|_{\mathrm{free}}
\]
\[
+\left.\frac{\hbar^4}{2m^2W}\int\!d^2\vek{x}\,
\nabla\psi^+(\vek{x},0)\cdot\nabla\psi(\vek{x},0)\right|_{\mathrm{bound}}
+\frac{W}{2}\int\!d^2\vek{x}\,\psi^+(\vek{x},0)\cdot\psi(\vek{x},0).
\]
This includes a superposition of kinetic terms similar to
$H_0$ (\ref{eq:ham0}) with
\[
\mu=\frac{m^2W}{\hbar^2}=\frac{m}{L_\perp},
\]
where $L_\perp=\kappa^{-1}$ is the bulk penetration depth of the bound 
states $|\vek{k},\kappa,+\rangle$.\\
Superposition of two-dimensional and three-dimensional kinetic terms
might appear like an exotic concept for the description of particle 
propagation in the presence of surfaces or interfaces, but on
a qualitative level the concept can be justified. And we have already
seen in equations (\ref{eq:g03}) and (\ref{eq:g02}) that it allows us to 
provide simple analytic estimates on the impact of dimensional 
competition on propagation effects. In the following we will focus on a 
study of the competition between the different kinetic terms, i.e. we 
will neglect any potential terms and study the Hamiltonian $H_0$.

\section{The energy dependent Green's function}\label{sec:green}

The energy dependent Green's function $\mathcal{G}(E)$
satisfies
\[
(E-H)\mathcal{G}(E)=1.
\]
The Green's function $G(E)$ which has (\ref{eq:g03}) as the
$E=0,z'=0$ limit is related to $\mathcal{G}(E)$ through
\[
\mathcal{G}(E)=-\frac{2m}{\hbar^2}G(E).
\]
Our objective is to find the energy dependent Green's function
$G(E)$ for the dimensionally hybrid Hamiltonian $H_0$ in equation 
(\ref{eq:ham0}). For later comparison, we will first revisit the free 
Green's function $G_0(E)$ of the Hamiltonian $H_0$ without the interface
term, $1/\mu\to 0$, and then calculate $G(E)$.

\subsection{The free Green's function}\label{sec:green0}

The free retarded Green's function is translation invariant,
\[
\langle\vek{x},z|G_0(E)|\vek{x}',z'\rangle
=G_0(E;\vek{x}-\vek{x}',z-z')
\]
and is well known to be
\begin{eqnarray}\label{eq:G0E}
G_0(E;\vek{x},z)&=&
\frac{\Theta(-E)}{4\pi\sqrt{r^2+z^2}}
\exp\!\left(-\frac{1}{\hbar}\sqrt{-2mE(r^2+z^2)}\right)
\\ \nonumber
&+&\frac{\Theta(E)}{4\pi\sqrt{r^2+z^2}}
\exp\!\left(\frac{\mathrm{i}}{\hbar}\sqrt{2mE(r^2+z^2)}\right),
\end{eqnarray}
where we continue to use cylinder coordinates. For positive
energy this corresponds to the standard choice of poles in 
$G_0(E;\vek{k},k_\perp)$ to generate outgoing spherical waves
without any incoming spherical component, in symbolic notation
\begin{eqnarray}\label{eq:G0Eret}
\mathcal{G}(E)&=&\frac{1}{E-H+\mathrm{i}\epsilon}
=\sum_{n,\nu}\frac{|n,\nu\rangle\langle n,\nu|}{E-E_n+\mathrm{i}\epsilon}
\\ \nonumber
&=&\mathcal{P}\sum_{n,\nu}\frac{|n,\nu\rangle\langle n,\nu|
}{E-E_n}-\mathrm{i}\pi\sum_{n,\nu}
\delta(E-E_n)|n,\nu\rangle\langle n,\nu|,
\end{eqnarray}
where $\nu$ is a degeneracy index.\\
Green's functions for surfaces or interfaces are commonly parametrized in 
an axially symmetric mixed representation like $G(E;\vek{k},z,z')$, see 
e.g. \cite{LF}. In bra-ket notation this corresponds for the free Green's 
function $G_0(E)$, which is also translation invariant in $z$ direction, 
to
\[
\langle\vek{k},z|G_0(E)|\vek{k}',z'\rangle
=G_0(E;\vek{k},z-z')\delta(\vek{k}-\vek{k}').
\]
 For later comparison we will also briefly recall the explicit form 
of the free Green's function $G_0(E)$ in the axially symmetric mixed
parametrization. The equation
\[
\left(\partial_z^2-\vek{k}^2+\frac{2mE}{\hbar^2}\right)G_0(E;\vek{k},z)
=-\delta(z)
\]
yields
\begin{eqnarray}\label{eq:g03axial}
G_0(E;\vek{k},z)&=&\frac{1}{2\pi}\int\!dk_\perp\,
\frac{\exp(\mathrm{i}k_\perp z)}{k^2_\perp+\vek{k}^2-(2mE/\hbar^2)
-\mathrm{i}\epsilon}
\\ \nonumber
&=&\frac{\hbar\Theta(\hbar^2\vek{k}^2-2mE)
}{2\sqrt{\hbar^2\vek{k}^2-2mE}}
\exp\!\left(-\sqrt{\hbar^2\vek{k}^2-2mE}\frac{|z|}{\hbar}\right)
\\ \nonumber
&+&\frac{\mathrm{i}\hbar\Theta(2mE-\hbar^2\vek{k}^2)
}{2\sqrt{2mE-\hbar^2\vek{k}^2}}
\exp\!\left(\mathrm{i}\sqrt{2mE-\hbar^2\vek{k}^2}
\frac{|z|}{\hbar}\right).
\end{eqnarray}
Note that this axially symmetric representation of the free Green's
function depends on the number $d-1$ of dimensions perpendicular to $z$ 
only through the number of components of the wavevector $\vek{k}$.

\subsection{The interface Green's function $G(E)$}\label{sec:GE}

Our objective is to generalize the Green's function
$G(\vek{x}-\vek{x}',z)=\langle\vek{x},z|G(0)|\vek{x}',0\rangle$
(\ref{eq:g03}) both to general values of $E$ and $z'$,
$\langle\vek{x},z|G(E)|\vek{x}',z'\rangle$. We will also
keep the general value $z_0$ for the location of the interface and
investigate the dependence of $G(E)$ on this parameter.\\
The Hamiltonian
\[
H_0=\int\!d^2\vek{x}\int\!dz\left(\frac{\hbar^2}{2m}
\left(\nabla\psi^+\cdot\nabla\psi
+\partial_z\psi^+\cdot\partial_z\psi\right)
+\frac{\hbar^2}{2\mu}\delta(z-z_0)\nabla\psi^+\cdot
\nabla\psi\right)
\]
yields the Schr\"odinger equation
\[
E\psi(\vek{x},z)=-\frac{\hbar^2}{2m}(\Delta+\partial_z^2)\psi(\vek{x},z)
-\frac{\hbar^2}{2\mu}\delta(z-z_0)\Delta\psi(\vek{x},z),
\]
or
\[
\left(E-\frac{\vek{p}^2+p_z^2}{2m}
-|z_0\rangle\langle z_0|\frac{\vek{p}^2}{2\mu}\right)|\psi\rangle=0,
\quad
\left(E-\frac{\vek{p}^2+p_z^2}{2m}
-|z_0\rangle\langle z_0|\frac{\vek{p}^2}{2\mu}\right)
\mathcal{G}(E)=1.
\]
The last equation in $(\vek{x},z)$ representation is
\begin{equation}\label{eq:trans1}
\left(\frac{2m}{\hbar^2}E+\Delta+\partial_z^2
+\delta(z-z_0)\frac{m}{\mu}\Delta\right)
\langle\vek{x},z|G(E)|\vek{x}',z'\rangle
=-\delta(\vek{x}-\vek{x}')\delta(z-z').
\end{equation}
Substitution of the Fourier transform
\[
\langle\vek{x},z|G(E)|\vek{x}',z'\rangle
=\frac{1}{4\pi^2}\int\!d^2\vek{k}\int\!d^2\vek{k}'\,
\langle\vek{k},z|G(E)|\vek{k}',z'\rangle
\exp[\mathrm{i}(\vek{k}\cdot\vek{x}-\vek{k}'\cdot\vek{x}')]
\]
yields
\begin{eqnarray}\label{eq:trans2}
&&\left(\frac{2m}{\hbar^2}E-\vek{k}^2+\partial_z^2\right)
\langle\vek{k},z|G(E)|\vek{k}',z'\rangle
-\frac{m}{\mu}\vek{k}^2\delta(z-z_0)
\langle\vek{k},z|G(E)|\vek{k}',z'\rangle
\\ \nonumber
&&=-\delta(\vek{k}-\vek{k}')\delta(z-z').
\end{eqnarray}
This yields with
\[
\langle\vek{k},z|G(E)|\vek{k}',z'\rangle
=\langle z|G(E,\vek{k})|z'\rangle\delta(\vek{k}-\vek{k}')
\]
the condition
\[
\left(\frac{2m}{\hbar^2}E-\vek{k}^2+\partial_z^2\right)
\langle z|G(E,\vek{k})|z'\rangle
-\frac{m}{\mu}\vek{k}^2\delta(z-z_0)
\langle z|G(E,\vek{k})|z'\rangle=-\delta(z-z').
\]
 Fourier transformation with respect to $z$ yields 
\begin{eqnarray}\label{eq:trans3}
&&\left(\frac{2m}{\hbar^2}E-\vek{k}^2-k^2_\perp\right)
\langle k_\perp|G(E,\vek{k})|z'\rangle
\\ \nonumber
&&-\frac{m}{2\pi\mu}\vek{k}^2\int\!d\kappa_\perp\,
\exp[\mathrm{i}(\kappa_\perp-k_\perp)z_0]
\langle\kappa_\perp|G(E,\vek{k})|z'\rangle
=-\frac{1}{\sqrt{2\pi}}\exp(-\mathrm{i}k_\perp z').
\end{eqnarray}
This result implies that 
$\langle k_\perp|G(E,\vek{k})|z'\rangle$ has the form
\[
\exp(\mathrm{i}k_\perp z_0)\langle k_\perp|G(E,\vek{k})|z'\rangle
=\frac{(\exp[\mathrm{i}k_\perp(z_0-z')]/\sqrt{2\pi})
+f(E,\vek{k},z')}{k^2_\perp+\vek{k}^2-(2mE/\hbar^2)}
\]
with the yet to be determined function $f(E,\vek{k},z')$
satisfying
\[
f(E,\vek{k},z')+\frac{m}{2\pi\mu}\vek{k}^2\int\!d\kappa_\perp\,
\frac{(\exp[\mathrm{i}\kappa_\perp(z_0-z')]/\sqrt{2\pi})
+f(E,\vek{k},z')}{\kappa^2_\perp+\vek{k}^2-(2mE/\hbar^2)}=0.
\]
 For the treatment of the integrals we should be consistent
with the calculation of the free retarded Green's function
(\ref{eq:g03axial}),
\begin{eqnarray*}
\int\!\frac{d\kappa_\perp}{2\pi}\,
\frac{\exp\!\left(\mathrm{i}\kappa_\perp z\right)}{
\kappa^2_\perp+\vek{k}^2-(2mE/\hbar^2)-\mathrm{i}\epsilon}&=&
\frac{\hbar}{2}\Theta(\hbar^2\vek{k}^2-2mE)
\frac{\exp\!\left(-\sqrt{\hbar^2\vek{k}^2-2mE}|z|/\hbar\right)}{
\sqrt{\hbar^2\vek{k}^2-2mE}}
\\
&+&\mathrm{i}\frac{\hbar}{2}\Theta(2mE-\hbar^2\vek{k}^2)
\frac{\exp\!\left(\mathrm{i}\sqrt{2mE-\hbar^2\vek{k}^2}|z|/\hbar\right)}{
\sqrt{2mE-\hbar^2\vek{k}^2}}.
\end{eqnarray*}
This yields
\begin{eqnarray*}
\Bigg[1&+&\frac{m\hbar}{2\mu}\vek{k}^2\left(
\frac{\Theta(\hbar^2\vek{k}^2-2mE)}{\sqrt{\hbar^2\vek{k}^2-2mE}}
+\mathrm{i}\frac{\Theta(2mE-\hbar^2\vek{k}^2)}{
\sqrt{2mE-\hbar^2\vek{k}^2}}\right)\Bigg]f(E,\vek{k},z')
\\
=&-&\frac{m\hbar}{2\mu\sqrt{2\pi}}\vek{k}^2\Bigg[
\frac{\Theta(\hbar^2\vek{k}^2-2mE)}{\sqrt{\hbar^2\vek{k}^2-2mE}}
\exp\!\left(-\sqrt{\hbar^2\vek{k}^2-2mE}\frac{|z'-z_0|}{\hbar}\right)
\\
&+&\mathrm{i}\frac{\Theta(2mE-\hbar^2\vek{k}^2)}{\sqrt{2mE-\hbar^2\vek{k}^2}}
\exp\!\left(\mathrm{i}\sqrt{2mE-\hbar^2\vek{k}^2}\frac{|z'-z_0|}{\hbar}\right)\Bigg],
\end{eqnarray*}
and therefore
\begin{eqnarray}\label{eq:fullg1}
\langle k_\perp|G(E,\vek{k})|z'\rangle
&=&\frac{1}{\sqrt{2\pi}}\frac{1}{k^2_\perp+\vek{k}^2-(2mE/\hbar^2)}
\Bigg[\exp(-\mathrm{i}k_\perp z')
\\ \nonumber
&-&\frac{\hbar\vek{k}^2\ell\Theta(\hbar^2\vek{k}^2-2mE)}{
\sqrt{\hbar^2\vek{k}^2-2mE}
+\hbar\vek{k}^2\ell}
\exp\!\left(-\mathrm{i}k_\perp z_0-\sqrt{\hbar^2\vek{k}^2-2mE}\frac{|z'-z_0|}{\hbar}\right)
\\ \nonumber
&-&\mathrm{i}\frac{\hbar\vek{k}^2\ell\Theta(2mE-\hbar^2\vek{k}^2)}{
\sqrt{2mE-\hbar^2\vek{k}^2}+\mathrm{i}\hbar\vek{k}^2\ell}
\exp\!\left(-\mathrm{i}k_\perp z_0+\mathrm{i}\sqrt{2mE-\hbar^2\vek{k}^2}\frac{|z'-z_0|}{\hbar}\right)
\Bigg].
\end{eqnarray}
Here we use the definition\footnote{
In the simple model from Section \ref{sec:ham}, 
$\ell=L_\perp/2=(2\kappa)^{-1}$ would be the bulk penetration depth of 
the probability densities $|\langle\vek{x},z|\vek{k},\kappa,+\rangle|^2$ 
of the bound states. But note that we have neglected any potential
contribution in $H_0$, such that the results derived here are not directly
applicable to the model from Section \ref{sec:ham}.}
\[
\ell\equiv\frac{m}{2\mu}.
\]
 Fourier transformation with respect to $k_\perp$ yields finally
\begin{eqnarray}\label{eq:fullg2}
\langle z|G(E,\vek{k})|z'\rangle
&=&\frac{\hbar\Theta(\hbar^2\vek{k}^2-2mE)
}{2\sqrt{\hbar^2\vek{k}^2-2mE}}\Bigg[
\exp\!\left(-\sqrt{\hbar^2\vek{k}^2-2mE}\frac{|z-z'|}{\hbar}\right)
\\ \nonumber
&-&\frac{\hbar\vek{k}^2\ell}{
\sqrt{\hbar^2\vek{k}^2-2mE}
+\hbar\vek{k}^2\ell}
\exp\!\left(-\sqrt{\hbar^2\vek{k}^2-2mE}\frac{|z-z_0|+|z'-z_0|}{\hbar}\right)\Bigg]
\\ \nonumber
&+&\mathrm{i}\frac{\hbar\Theta(2mE-\hbar^2\vek{k}^2)
}{2\sqrt{2mE-\hbar^2\vek{k}^2}}\Bigg[
\exp\!\left(\mathrm{i}\sqrt{2mE-\hbar^2\vek{k}^2}\frac{|z-z'|}{\hbar}\right)
\\ \nonumber
&-&\mathrm{i}\frac{\hbar\vek{k}^2\ell}{
\sqrt{2mE-\hbar^2\vek{k}^2}+\mathrm{i}\hbar\vek{k}^2\ell}
\exp\!\left(\mathrm{i}\sqrt{2mE-\hbar^2\vek{k}^2}\frac{|z-z_0|+|z'-z_0|}{\hbar}\right)\Bigg].
\end{eqnarray}

This result is translation invariant in the transverse $z$ direction
for scattering off perturbations on the interface, $z'=z_0$,
\[
\langle z|G(E,\vek{k})|z'\rangle\Big|_{z'=z_0}=G(E,\vek{k},z-z_0).
\]

\section{Density of states and Fermi energy on the interface
in the dimensionally hybrid model}\label{sec:rho}


The equation (\ref{eq:G0Eret}) yields a standard expression for the
density of states in terms of the imaginary part of Green's functions, 
see e.g. \cite{DS},
\begin{eqnarray}\label{eq:rho0E}
\varrho(E_n,\vek{x},z)
&=&g\sum_{\nu}\langle\vek{x},z|n,\nu\rangle\langle n,\nu|\vek{x},z\rangle
=-\frac{g}{\pi}
\Im\langle\vek{x},z|\mathcal{G}(E_n)|\vek{x},z\rangle
\\ \nonumber
&=&\frac{2mg}{\pi\hbar^2}\Im\langle\vek{x},z|G(E_n)|\vek{x},z\rangle.
\end{eqnarray}
Here we explicitly included a factor $g$ for the number of spin or
helicity states, because the summation over degeneracy indices in
(\ref{eq:G0Eret}) usually only involves orbital indices.\\
 For translation invariant Green's functions
\[
\langle\vek{x},z|\mathcal{G}(E)|\vek{x}',z'\rangle
=\mathcal{G}(E;\vek{x}-\vek{x}',z-z')
\]
we have
\begin{equation}\label{eq:rhoE1}
\varrho(E)=-\frac{g}{\pi}\Im\mathcal{G}(E;\vek{x}=0,z=0)
=-\frac{g}{\pi(2\pi)^{d-1}}\Im\int\!d^{d-1}\vek{k}\,
\mathcal{G}(E;\vek{k},z=0).
\end{equation}
Insertion of the free retarded propagator (\ref{eq:g03axial}) reproduces 
the standard density of states (\ref{eq:rhod}), of course,
\begin{eqnarray*}
\varrho_{(d)}(E)&=&\frac{2mg}{\pi\hbar^2}
\frac{\hbar\Theta(E)}{2^{d-1}\sqrt{\pi}^{d-1}\Gamma((d-1)/2)}
\int_0^{\sqrt{2mE}/\hbar}\!dk\,\frac{k^{d-2}}{\sqrt{2mE-\hbar^2k^2}}
\\
&=&g\Theta(E)\sqrt{\frac{m}{2\pi}}^d
\frac{\sqrt{E}^{d-2}}{\Gamma(d/2)\hbar^d}.
\end{eqnarray*}

The interface at $z_0$ breaks translational invariance in $z$ direction,
and we have with $\langle\vek{k},z|G(E)|\vek{k}',z\rangle=
\langle z|G(E,\vek{k})|z\rangle\delta(\vek{k}-\vek{k}')$
\[
\varrho(E,z)=
\frac{2mg}{\pi\hbar^2}\Im\langle\vek{x},z|G(E)|\vek{x},z\rangle
=\frac{2mg}{\pi\hbar^2}\Im\int\!\frac{d^2\vek{k}}{4\pi^2}\,
\langle z|G(E,\vek{k})|z\rangle.
\]

We will use the result (\ref{eq:fullg2}) to calculate the density of 
states $\varrho(E,z_0)$ for the Hamiltonian (\ref{eq:ham0}) on the
interface. Substitution yields
\begin{eqnarray*}
\varrho(E,z_0)&=&
\frac{gm}{2\pi^3\hbar^2}\Im\int\!d^2\vek{k}\,
\langle z_0|G(E,\vek{k}|z_0\rangle
\\
&=&\frac{gm}{2\pi^2\hbar}\Theta(E)\int_0^{\sqrt{2mE}/\hbar}
\!dk\,k\frac{\sqrt{2mE-\hbar^2k^2}}{2mE-\hbar^2k^2+\hbar^2k^4\ell^2}.
\end{eqnarray*}
The evaluation of the integral yields
\begin{eqnarray}\label{eq:rhointer}
\varrho(E,z_0)&=&\frac{gm\Theta(E)}{
4\pi^2\hbar^2\ell\sqrt{\hbar^2-8mE\ell^2}}\Theta(\hbar^2-8mE\ell^2)
\\ \nonumber
&\times&
\left[\left(\hbar+\sqrt{\hbar^2-8mE\ell^2}\right)
\mathrm{\arctan}\!\left(\frac{\ell\sqrt{8mE}}{
\hbar+\sqrt{\hbar^2-8mE\ell^2}}\right)\right.
\\ \nonumber
&-&\left.\left(\hbar-\sqrt{\hbar^2-8mE\ell^2}\right)
\mathrm{\arctan}\!\left(\frac{\ell\sqrt{8mE}}{
\hbar-\sqrt{\hbar^2-8mE\ell^2}}\right)
\right]
\\ \nonumber
&+&\frac{gm\Theta(8mE\ell^2-\hbar^2)}{4\pi^2\hbar^2\ell}
\left[\frac{\hbar}{\sqrt{8mE\ell^2-\hbar^2}}
\ln\!\left(\frac{\ell\sqrt{8mE}-\sqrt{8mE\ell^2-\hbar^2}}{\hbar}\right)
\right.
\\ \nonumber
&+&\left.\mathrm{arctan}\!\left(
\frac{\sqrt{8mE\ell^2-\hbar^2}+\ell\sqrt{8mE}}{\hbar}\right)
+\mathrm{arctan}\!\left(
\frac{\ell\sqrt{8mE}-\sqrt{8mE\ell^2-\hbar^2}}{\hbar}\right)
\right].
\end{eqnarray}
This is a more complicated result than the
density (\ref{eq:rhod}) for $d=2$ or $d=3$. However, it 
reduces to either the two-dimensional or three-dimensional
density of states in the appropriate limits.
 For large energies, i.e. if the states only probe length scales
smaller than the transverse penetration depth $\ell$,
we find the two-dimensional density of states properly rescaled by
a dimensional factor to reflect that it is a density of states
per three-dimensional volume,
\begin{equation}\label{eq:rho2}
8mE\ell^2\gg\hbar^2:\quad
\varrho(E,z_0)\to\Theta(E)\frac{gm}{8\pi\hbar^2\ell}=\frac{1}{4\ell}
\varrho_{(d=2)}(E).
\end{equation}


\begin{center}
\begin{figure}[t]
\hspace*{50mm}\scalebox{0.4}{\includegraphics{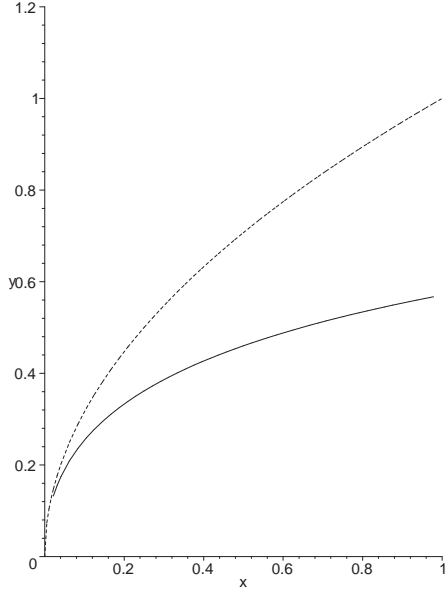}}
\caption{\label{fig:223rho1}
The dotted line is the three-dimensional density of states in
units of $gm/4\pi^2\hbar^2\ell$.
The continuous line is the density of states
(\ref{eq:rhointer}). $0\le x=8mE\ell^2/\hbar^2\le 1$.}
\end{figure}
\end{center}

 For small energies, i.e. if the states probe length scales
larger than the transverse penetration depth $\ell$,
we find the three-dimensional density of states
\begin{equation}\label{eq:rho3}
8mE\ell^2\ll\hbar^2:\quad\varrho(E,z_0)\to\Theta(E)\frac{g\sqrt{m}^3}{
\sqrt{2}\pi^2\hbar^3}\sqrt{E}=\varrho_{(d=3)}(E).
\end{equation}
This limiting behavior for interpolation between two and three
dimensions is consistent with what was observed already for the
zero energy Green's function in the interface 
(\ref{eq:g02}-\ref{eq:rggell}).\\
Equation (\ref{eq:rhointer}) also implies interpolating behavior
for the relation between electron density and Fermi energy on the
interface for the Hamiltonian (\ref{eq:ham0}). The full relation
is with $g=2$ 
\begin{eqnarray}\nonumber
n(z_0)&=&\int_0^{E_F}\!dE\,\varrho(E,z_0)
\\ \label{eq:ninter}
&=&
\frac{\sqrt{mE_F}}{\sqrt{8}\pi^2\hbar\ell^2}-\frac{1}{16\pi\ell^3}
\\ \nonumber
&+&\frac{\Theta(\hbar^2-8mE_F\ell^2)}{8\pi^2\hbar^2\ell^3}
\Bigg[\left(4mE_F\ell^2-
\hbar\sqrt{\hbar^2-8mE_F\ell^2}\right)
\mathrm{arctan}\!\left(\frac{
\sqrt{8mE_F}\ell}{\hbar+\sqrt{\hbar^2-8mE_F\ell^2}}\right)
\\ \nonumber
&+&\left(4mE_F\ell^2+
\hbar\sqrt{\hbar^2-8mE_F\ell^2}\right)
\mathrm{arctan}\!\left(\frac{
\sqrt{8mE_F}\ell}{\hbar-\sqrt{\hbar^2-8mE_F\ell^2}}\right)
\Bigg]
\\ \nonumber
&+&\frac{\Theta(8mE_F\ell^2-\hbar^2)}{8\pi^2\hbar\ell^3}
\Bigg[\sqrt{8mE_F\ell^2-\hbar^2}\ln\!\left(
\frac{\sqrt{8mE_F}\ell-\sqrt{8mE_F\ell^2-\hbar^2}}{\hbar}\right)
\\ \nonumber
&+&\frac{4mE_F\ell^2}{\hbar}
\mathrm{arctan}\!\left(
\frac{\sqrt{8mE_F}\ell+\sqrt{8mE_F\ell^2-\hbar^2}}{\hbar}\right)
\\ \nonumber
&+&\frac{4mE_F\ell^2}{\hbar}
\mathrm{arctan}\!\left(
\frac{\sqrt{8mE_F}\ell-\sqrt{8mE_F\ell^2-\hbar^2}}{\hbar}
\right)\Bigg].
\end{eqnarray}
This approximates two-dimensional behavior for $mE_F\ell^2\gg\hbar^2$,
\[
n(z_0)\simeq\frac{mE_F}{4\pi\hbar^2\ell}=\frac{1}{4\ell}n_{(d=2)},
\]
and three-dimensional behavior for $mE_F\ell^2\ll\hbar^2$,
\[
n(z_0)\simeq\frac{\sqrt{2mE_F}^3}{3\pi^2\hbar^3}=n_{(d=3)}.
\]
It is intuitively understandable that the presence of a layer
increases the Fermi energy for a given density of electrons.
The presence of a layer implies boundary or matching conditions
which reduce the number of available states at a given energy.

\section{Conclusion and Outlook}\label{sec:conc}

Inclusion of competing kinetic terms for effective propagation
of particles in low-dimensional subsystems and bulk
materials implies dimensional interpolation effects for
Green's functions and for quantities derived from the Green's
functions in such systems. Simple model systems with competing
kinetic terms can be solved exactly and yield analytic insights
into the transition between two-dimensional and three-dimensional
properties in materials with low-dimensional subsystems.\\
The present paper focused on the study of the competition between 
two-dimensional and three-dimensional kinetic terms, thereby neglecting 
any bulk and layer potentials. In that case the transition scale between 
two-dimensional and three-dimensional behavior is given by 
$\ell\sim m/\mu\sim (m/m_\ast)L$. Please note that inclusion of the 
confining layer potentials will change this scale for most systems. 
The approximation $H_0$ {\it per se} as an approximation for specific
systems is only useful for the study of propagation effects and impurity
scattering of particles which are not strongly bound to a thin layer,
but which are affected by its presence to the extent that propagation
in the thin layer is described by an effective mass.\\
The purpose of the present investigation was to further advance a novel 
tool for the study of low-dimensional systems, not to derive generic 
quantitative properties of these systems. The Green's function $G(E)$ 
calculated here should nevertheless prove useful for the study of 
impurity scattering of weakly coupled particles in thin layer systems, 
with the impurity potentials treated as perturbations.\\[2mm]

{\bf Acknowledgement:} This work was supported in part
by NSERC Canada.


\begin{thebibliography}{88}

\bibitem{rdtp03}
R. Dick, Int. J. Theor. Phys. {\bf 42} (2003) 569-581,
cond-mat/0204534.

\bibitem{rbl}
R.B. Laughlin, Phys. Rev. Lett. {\bf 50} (1983) 1395-1398.

\bibitem{TP} 
T. Chakraborty, P. Pietil\"ainen,
{\it The Quantum Hall Effects}, 2nd ed.,
Springer-Verlag, Berlin 1995.

\bibitem{tapash}
T. Chakraborty, Adv. Phys. {\bf 49} (2000) 959-1014.

\bibitem{patil}
S.H. Patil, Am. J. Phys. {\bf 68} (2000) 712-714.

\bibitem{LF} M. Lannoo, P. Friedel,
{\it Atomic and Electronic Structure of Surfaces},
Springer-Verlag, Berlin 1991.

\bibitem{DS} S. Doniach, E.H. Sondheimer,
{\it Green's Functions for Solid State Physicists},
Imperial College Press, London 1998.

\end{thebibliography}
\end{document}